\documentclass[12pt]{article}
\usepackage{epsfig}
\def\be{\begin{equation}}
\def\ee{\end{equation}}
\def\bea{\begin{eqnarray}}
\def\eea{\end{eqnarray}}
\usepackage{graphicx}

\catcode`\@=11
\def\lsim{\mathrel{\mathpalette\@versim<}}
\def\gsim{\mathrel{\mathpalette\@versim>}}
\def\@versim#1#2{\vcenter{\offinterlineskip
\ialign{$\m@th#1\hfil##\hfil$\crcr#2\crcr\sim\crcr } }}
\catcode`\@=12
\usepackage{axodraw}

\catcode`@=12
\begin{document}
\thispagestyle{empty}
\begin{flushright}
UCRHEP-T522\\
August 2012\
\end{flushright}
\vspace{0.8in}
\begin{center}
{\LARGE \bf Radiative Scaling Neutrino Mass\\ 
and Warm Dark Matter\\}
\vspace{1.2in}
{\bf Ernest Ma\\}
\vspace{0.2in}
{\sl Department of Physics and Astronomy, University of California,\\
Riverside, California 92521, USA\\}
\end{center}
\vspace{1.2in}
\begin{abstract}\
A new and radical scenario of the simple 2006 model of radiative neutrino 
mass is proposed, where there is no seesaw mechanism, i.e. neutrino masses are 
not inversely proportional to some large mass scale, contrary to the prevalent 
theoretical thinking.  The neutral singlet fermions in the loop have masses 
of order 10 keV, the lightest of which is absolutely stable and the others 
are very long-lived.  All are components of warm dark matter, which is 
a possible new paradigm for explaining the structure of the Universe at 
all scales.

\end{abstract}

\newpage
\baselineskip 24pt

Neutrino mass and dark matter are two important issues in physics and 
astrophysics.  In 2006, it was proposed~\cite{m06} that they are in fact 
intimately related in a model of radiative Majorana neutrinio mass 
in one loop, where the particles appearing in the loop are odd under an 
exactly conserved $Z_2$ symmetry~\cite{dm78}, thus allowing them to be 
dark-matter 
candidates.  The model is simplicity itself.  It extends the minimal 
standard model of particle interactions to include three neutral singlet 
fermions $N_k$ and a second scalar doublet $\eta = (\eta^+,\eta^0)$, 
all of which are odd under the aforementioned $Z_2$, whereas all 
standard-model particles are even.  Thus the observed neutrinos $\nu_i$ 
are forbidden to couple to $N_k$ through the standard-model Higgs doublet 
$\Phi = (\phi^+,\phi^0)$, but the new interactions $h_{ik} (\nu_i \eta^0 - 
l_i \eta^+) N_k + H.c.$ as well as the mass terms $(M_k/2) N_k N_k + H.c.$ 
are allowed.   Hence Majorana neutrino masses are generated in one loop 
as shown in Fig.~1.
\begin{figure}[htb]
\begin{center}
\begin{picture}(260,120)(0,0)
\ArrowLine(60,10)(90,10)
\ArrowLine(130,10)(90,10)
\ArrowLine(130,10)(170,10)
\ArrowLine(200,10)(170,10)
\DashArrowArc(130,10)(40,90,180)5
\DashArrowArcn(130,10)(40,90,0)5
\DashArrowLine(100,80)(130,50)5
\DashArrowLine(160,80)(130,50)5
\Text(75,0)[]{\large $\nu$}
\Text(185,0)[]{\large $\nu$}
\Text(110,0)[]{\large $N$}
\Text(150,0)[]{\large $N$}
\Text(100,52)[]{\large $\eta^0$}
\Text(160,52)[]{\large $\eta^0$}
\Text(95,90)[]{\large $\langle \phi^0 \rangle$}
\Text(165,90)[]{\large $\langle \phi^0 \rangle$}
\Text(130,10)[]{\Large $\times$}
\end{picture}
\end{center}
\caption{One-loop generation of scotogenic Majorana neutrino mass.}
\end{figure}
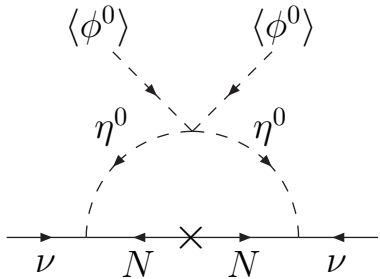
This mechanism has been called ``scotogenic'', from the Greek ``scotos'' 
meaning darkness.  Because of the allowed $(\lambda_5/2) (\Phi^\dagger \eta)^2 
+ H.c.$ interaction, $\eta^0 = (\eta_R + i \eta_I)/\sqrt{2}$ is split so 
that $m_R \neq m_I$.  The diagram of Fig.~1 can be computed exactly~\cite{m06}, 
i.e.
\begin{equation}
({\cal M}_\nu)_{ij} = \sum_k {h_{ik} h_{jk} M_k \over 16 \pi^2} 
\left[ {m_R^2 \over m_R^2 - M_k^2} \ln {m_R^2 \over M_k^2} - 
{m_I^2 \over m_I^2 - M_k^2} \ln {m_I^2 \over M_k^2} \right].
\end{equation} 
A good dark-matter candidate is $\eta_R$ as first pointed out in 
Ref.~\cite{m06}.  It was subsequently proposed by itself in Ref.~\cite{bhr06} 
and studied in detail in Ref.~\cite{lnot07}.  The $\eta$ doublet has become 
known as the ``inert'' Higgs doublet, but it does have gauge and scalar 
interactions even if it is the sole addition to the standard model. 
It may enable the scalar sector of this model to have a 
strong electroweak phase transition~\cite{cnsz12,bc12,gck12}, thus allowing 
for successful baryogenesis.  Because of the $\lambda_3 (\Phi^\dagger \Phi)
(\eta^\dagger \eta)$ interaction, it may also contribute significantly to 
$H \to \gamma \gamma$~\cite{p11,abg12}.

The usual assumption for neutrino mass in Eq.~(1) is
\begin{equation}
m_I^2 - m_R^2 << m_I^2 + m_R^2 << M_k^2,
\end{equation}
in which case
\begin{equation}
({\cal M}_\nu)_{ij} = {\lambda_5 v^2 \over 8 \pi^2} \sum_k 
{h_{ik} h_{jk} \over M_k} \left[ \ln {M_k^2 \over m_0^2} - 1 \right],
\end{equation}
where $m_0^2 = (m_I^2 + m_R^2)/2$ and $m_R^2 - m_I^2 = 2 \lambda_5 v^2$ 
($v = \langle \phi^0 \rangle$).  This scenario is often referred to as 
the radiative seesaw.  There is however another very interesting scenario, 
i.e. 
\begin{equation}
M_k^2 << m_R^2, ~m_I^2.
\end{equation}
Neutrino masses are then given by
\begin{equation}
({\cal M}_\nu)_{ij} = {\ln (m_R^2/m_I^2) \over 16 \pi^2} \sum_k 
h_{ik} h_{jk} M_k.
\end{equation}
This simple expression is actually very extraordinary, because the 
prevalent theoretical thinking on neutrino masses is that they should be 
inversely proportional to some large mass scale, coming from the 
dimension-five operator~\cite{w79}
\begin{equation}
{\cal L} = {f_{ij} \over 2\Lambda}(\nu_i \phi^0 - l_i \phi^+)
(\nu_j \phi^0 - l_j \phi^+) + H.c.,
\end{equation}
whereas Eq.~(5) is clearly not of this form, unless of course 
$|m_R^2-m_I^2| << m^2_{R,I}$.  It also allows neutrino 
masses to be of order 0.1 eV and $M_k$ of order 10 keV, with $h_{ik}^2$ 
of order $10^{-3}$.  If $M_k = 0$, then $N_k$ may be assigned $L=-1$ and 
$L$ would be conserved. Thus $M_k \neq 0$ corresponds to the breaking of 
$L$ to $(-1)^L$ and $M_k$ could be naturally small, as compared to the 
electron mass which preserves $L$.  Without the $Z_2$ 
symmetry, the canonical seesaw mechanism~\cite{seesaw}, i.e. $m_\nu = 
-m_D^2/M$, would require the $\nu N$ Dirac mass $m_D$ to be of order 30 eV 
and the $\nu-N$ mixing, i.e. $m_D/M$, to be of order $3 \times 10^{-3}$.  
This would render $N$ unacceptable as a dark-matter candidate.  As it is, 
with $Z_2$ and $\eta$, there is no $m_D$ and no $\nu-N$ mixing.  There is 
also no seesaw.  Each neutrino mass is simply proportional to a linear 
combination of $M_k$ according to Eq.~(5).  Hence their ratio is just a 
scale factor and small neutrino masses are due to this ``scaling'' mechanism. 
Note that the interesting special case where only $M_1$ is small, with all 
the other masses of order $10^2$ GeV, has been considered 
previously~\cite{gop10}. 

If $\eta^\pm, \eta_R, \eta_I$ are of order $10^2$ GeV, the interactions of 
$N_k$ with the neutrinos and charged leptons are weaker than the usual 
weak interaction, hence $N_k$ may be considered ``sterile'' 
and become excellent candidates of warm dark matter, which is a 
possible new paradigm for explaining the structure of the Universe 
at all scales~\cite{dvs11,dvfs12}.
However, unlike the usual sterile neutrinos~\cite{white12} which mix with 
the active neutrinos, the lightest $N_k$ here is absolutely stable.  This 
removes one of the most stringent astrophysical constraints on warm dark 
matter, i.e. the absence of galactic X-ray emission from its decay, which 
would put an upper bound of perhaps 2.2 keV on its mass~\cite{wlp12}, whereas 
Lyman-$\alpha$ forest observations (which still apply in this case) impose 
a lower bound of perhaps 5.6 keV~\cite{vbbhrs08}.  Such a stable sterile 
neutrino (called a ``scotino'') 
was already discussed recently~\cite{m12} in a left-right extension of the 
standard model, but the present proposal is far simpler.  Conventional 
left-right models where the $SU(2)_R$ neutrinos mix with the $SU(2)_L$ 
neutrinos have also been studied~\cite{bhl10,bkl12,nsz12}.

The diagram of Fig.~1 is always accompanied by that of $l_i \to l_j \gamma$ 
as shown in Fig.~2.
\begin{figure}[htb]
\begin{center}
\begin{picture}(260,120)(0,0)
\ArrowLine(60,10)(90,10)
\ArrowLine(170,10)(90,10)
\ArrowLine(170,10)(200,10)
\DashArrowArc(130,10)(40,90,180)5
\DashArrowArc(130,10)(40,0,90)5
\Photon(130,50)(160,80)33
\Text(75,0)[]{\large $l_i$}
\Text(185,0)[]{\large $l_j$}
\Text(130,0)[]{\large $N$}
\Text(100,52)[]{\large $\eta^+$}
\Text(160,52)[]{\large $\eta^+$}
\Text(165,90)[]{\large $\gamma$}
\end{picture}
\end{center}
\caption{Radiative decay of $l_i \to l_j \gamma$.}
\end{figure}
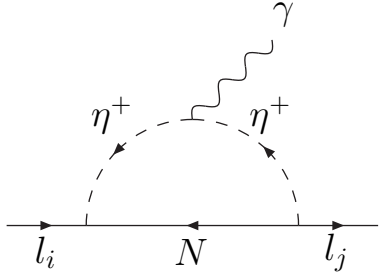
For $\mu \to e \gamma$, this branching fraction is given by~\cite{kms06}
\begin{equation}
B(\mu \to e \gamma) = {\alpha \over 768 \pi} {|\sum_k h_{\mu k} h^*_{e k}|^2 
\over (G_F m^2_{\eta^+})^2}.
\end{equation}
Using the experimental upper bound~\cite{meg11} of $2.4 \times 10^{-12}$, 
this implies
\begin{equation}
m_{\eta^+} > 310~{\rm GeV}~~(|\sum_k h_{\mu k} h^*_{e k}|/10^{-3})^{1/2}.
\end{equation}
Note that $m_{\eta^+}$ of order 300 GeV is possible  
for electroweak baryogenesis.  Together, this would imply that 
$B(\mu \to e \gamma)$ is required by this model to be just below the present 
bound. As for the muon anomalous magnetic moment, it is given by~\cite{mr01}
\begin{equation}
\Delta a_\mu = - {m_\mu^2 \over 96 \pi^2 m^2_{\eta^+}} \sum_k |h_{\mu k}|^2.
\end{equation}
Hence Eq.~(8) implies
\begin{equation}
|\Delta a_\mu| < 1.23 \times 10^{-13} ~~(\sum_k |h_{\mu k}|^2/|\sum_k 
h_{\mu k} h^*_{e k}|),
\end{equation}
much below the experimental uncertainty of $6 \times 10^{-10}$.

Since $N_k$ are assumed light, muon decay also proceeds at tree level 
through $\eta^+$ exchange, i.e. $\mu \to N_\mu e \bar{N}_e$.  The 
inclusive rate is easily calculated to be
\begin{equation}
\Gamma (\mu \to N_\mu e \bar{N}_e) = { (\sum_k |h_{\mu k}|^2)(\sum_k |h_{e k}|^2) 
m^5_{\mu} \over 6144 \pi^3 m^4_{\eta^+}}.
\end{equation}
Since $N_{\mu}$ and $\bar{N}_e$ are invisible just as $\nu_\mu$ and 
$\bar{\nu}_e$ are invisible in the dominant decay $\mu \to \nu_\mu e 
\bar{\nu}_e$ (with rate $G_F^2 m_\mu^5/192 \pi^3$), this would change 
the experimental value of $G_F$.  However, their ratio $R$ is very 
small.  Using Eq.~(8),
\begin{equation}
R < 2.5 \times 10^{-8} ~~(\sum_k |h_{\mu k}|^2) (\sum_k |h_{e k}|^2) / 
|\sum_k h_{\mu k} h^*_{e k}|^2,
\end{equation}
much below the experimental uncertainty of $10^{-5}$.

Whereas the lightest scotino, called it $N_1$, is absolutely stable, 
$N_{2,3}$ will decay into $N_1$ through $\eta_R$ and $\eta_I$.  The decay rate 
of $N_2 \to N_1 \bar{\nu}_i \nu_j$ is given by
\begin{eqnarray}
&& \Gamma (N_2 \to N_1 \bar{\nu}_i \nu_j) = {|h_{i2} h^*_{j1}|^2 \over 
256 \pi^3 M_2} \left( {1 \over m_R^2} + {1 \over m_I^2} \right)^2 \nonumber \\ 
&& \times \left( {M_2^6 \over 96} - {M_1^2 M_2^4 \over 12} + {M_1^6 \over 12} 
- {M_1^8 \over 96 M_2^2} + {M_1^4 M_2^2 \over 8} \ln {M_2^2 \over M_1^2} 
\right).
\end{eqnarray}
Let $M_2 - M_1 = \Delta M$ be small compared to $M_2$, then
\begin{equation}
\Gamma (N_2 \to N_1 \bar{\nu}_i \nu_j) = {|h_{i2} h^*_{j1}|^2 (\Delta M)^5 \over 
1920 \pi^3} \left( {1 \over m_R^2} + {1 \over m_I^2} \right)^2.
\end{equation}
As an example, let $\Delta M = 1$ keV, $|h_{i2} h^*_{j1}|^2 = 
10^{-6}$, $m_R = 240$ GeV, $m_I = 150$ GeV, then $\Gamma = 6.42 \times 
10^{-50}$ GeV, corresponding to a decay lifetime of $3.25 \times 10^{17}$ y, 
which is much longer than the age of the Universe, i.e. 
$13.75 \pm 0.11 \times 10^9$ y.  This means that $N_{1,2,3}$ may all be  
components of dark matter today.  Note that $N_2 \to N_1 \gamma$ 
is now possible with $E_\gamma \simeq \Delta M$, but since $\Delta M$ 
may be small, say 1 keV, whereas $M_{1,2,3} \sim 10$ keV, the tension 
between galactic X-ray data and Lyman-$\alpha$ forest observations 
is easily relaxed.  Note also that $\Gamma(N_2 \to N_1 \gamma)$ is 
proportional to $|\sum_i h_{i2} h^*_{i1}|^2$ which may be very much 
suppressed relative to $(\sum_i |h_{i2}|^2)(\sum_i |h_{i1}|^2) \sim 10^{-6}$.

The effective $N \bar{N} \to l \bar{l}, \nu \bar{\nu}$ interactions 
are of order $h^2/m^2_\eta \sim 10^{-8}$ GeV$^{-2}$, hence they remain 
in thermal equilibrium in the early Universe until a temperature of 
a few GeV.  Their number density $n_N$ is given by~\cite{bot01}
\begin{equation}
{n_N \over n_\gamma} = \left( {43/4 \over g^*_{dec}} \right) \left( 
{2 \over 11/2} \right) {3/2 \over 2},
\end{equation}
where $g^*_{dec} = 16$ in this model, counting $N_{1,2,3}$ in addition to 
photons, electrons, and the three neutrinos.  Their relic abundance at 
present would then be~\cite{bot01}
\begin{equation}
\Omega_N h^2 \simeq {115 \over 16} \left( {\sum_i M_i \over {\rm keV}} \right).
\end{equation}
For $\sum_i M_i \sim 30$ keV, this would be $1.9 \times 10^3$ times 
the measured value~\cite{wmap11} of $0.1123 \pm 0.0035$. 
The usual solution to this problem is to invoke a particle which 
decouples after $N_1$ and decays later as it becomes nonrelativistic, 
with a large release of entropy.  It is a well-known mechanism 
and has been elaborated recently~\cite{white12,bhl10,nsz12,km12,rt12} 
in the context of warm dark matter.

Another solution is to assume that the reheating temperature of the 
Universe is below a few GeV, so that $N_i$ are not thermally produced. 
Instead, they come from the decay of a scalar singlet $S$ with the 
allowed interaction $S N_i N_j$.  To accomplish this, consider 
$m_S = 2$ GeV.  Assume that $S$ decouples as it becomes 
nonrelativistic with an annihilation cross section times relative velocity 
of about $10^{-5}$ pb.  If $S$ is stable, this would correspond to a 
very large relic density; but $S$ decays to $NN$, so the actual relic 
density (i.e. that of $N$) is reduced by the factor $2 M/m_S \simeq 10^{-5}$.  
Since $\langle \sigma v \rangle$ is inversely proportional to relic density, 
this makes it effectively 1 pb, and yields the correct observed dark-matter 
relic density of the Universe.

The interactions of $S$ with itself and the particles of the scotogenic 
model are given by
\begin{eqnarray}
-{\cal L}_{int} &=& {1 \over 3} \mu_1 S^3 + \mu_2 S (\Phi^\dagger \Phi) 
+ \mu_3 S (\eta^\dagger \eta) + {1 \over 2} f_{ij} S N_i N_j + H.c. \nonumber \\
&+& {1 \over 4} \lambda_2 S^4 + {1 \over 2} \lambda_3 S^2 (\Phi^\dagger \Phi) 
+ {1 \over 2} \lambda_4 S^2 (\eta^\dagger \eta).
\end{eqnarray}
Assume $\mu_{1,2}$ to be negligible, so that $S-H$ mixing may be ignored, 
where $H$ is the physical Higgs boson with a mass of about 125 GeV. 
Assume $f_{ij} < 10^{-4}$, so that $N_i$ does not enter into thermal 
equilibrium through its interaction with $S$ below a few GeV. 
However, the interaction $\sqrt{2} \lambda_3 v H S^2$ will allow $S$ 
to thermalize because $H$ couples to quarks and leptons.  For $\lambda_3 
\sim 10^{-3}$, $\langle \sigma v \rangle \sim 10^{-5}$ pb may be obtained.

Consider now the phenomenology of this model at the Large Hadron Collider 
(LHC).  The decay rate of $H \to S S$ is given by
\begin{equation}
\Gamma (H \to S S) = {\lambda_3^2 v^2 \over 4 \pi m_H} \sim  
0.02 ~{\rm MeV},
\end{equation}
compared to the expected total width of about 4.3 MeV in the standard 
model.  Since $S$ decays into $NN$, this appears as an invisible 
decay of $H$.  However, this branching fraction is less than 0.5 percent, 
so it will be very difficult to check. As for the extra scalar particles 
$\eta^\pm, \eta_R, \eta_I$, they may be produced at the LHC through their 
electroweak gauge interactions.  Once produced, the decay
$\eta^+ \to l_i^+ N_j$ may be observed, but the decays 
$\eta_{R,I} \to \nu_i N_j$ are invisible.  If kinematically allowed, 
$\eta^+ \to \eta_{R,I} W^+$ and $\eta_R \to \eta_I Z$ are possible 
signatures.  The case $m_I > m_R$ with $m_I - m_R < m_Z$, so that 
$\eta_I \to \eta_R + {\rm virtual}~Z \to \eta_R + \mu^+ \mu^-$ has already 
been studied in some detail~\cite{cmr07}.

In conclusion, the scotogenic model~\cite{m06} of neutrino mass has been 
shown to admit a solution where there is no seesaw mechanism and $N_{1,2,3}$ 
may have masses of about 10 keV.  They are suitable as components of warm 
dark matter for explaining the structure of the Universe at all 
scales~\cite{dvs11,dvfs12}.  Since $N_1$ is absolutely stable, whereas 
$N_{2,3} \to N_1 \gamma$ is suppressed with $E_\gamma \simeq \Delta M$, 
the galactic X-ray upper bound of perhaps 2.2 keV on its mass~\cite{wlp12} 
is avoided.  It will also not be detected in terrestrial experiments.  
On the other hand, since this model requires an extra scalar doublet, 
it may be tested at the Large Hadron Collider.

\noindent \underline{Acknowledgment}~:~ This work is supported in part 
by the U.~S.~Department of Energy under Grant No.~DE-AC02-06CH11357.

\bibliographystyle{unsrt}

\end{document}